\documentclass{siamltex1213}

\usepackage{ifpdf}
\ifpdf
\else

\usepackage{graphicx}
\usepackage{float}
\usepackage{amsfonts}

\newcommand{\ben}{\begin{equation}}
\newcommand{\een}{\end{equation}}
\newcommand{\bea}{\begin{eqnarray}}
\newcommand{\eea}{\end{eqnarray}}
\newcommand{\ba}{\begin{array}}
\newcommand{\ea}{\end{array}}
\newcommand{\bit}{\begin{itemize}}
\newcommand{\eit}{\end{itemize}}

\renewcommand{\theequation}{\thesection.\arabic{equation}}
\renewcommand{\thefigure}{\thesection.\arabic{figure}}
\renewcommand{\thetable}{\thesection.\arabic{table}}

\def\math{\mathsurround 0pt}
\def\oversim#1#2{\lower.5pt\vbox{\baselineskip0pt \lineskip-.5pt
        \ialign{$\math#1\hfil##\hfil$\crcr#2\crcr{\scriptstyle\sim}\crcr}}}

\newcommand\lf{{\sc lat}field} 
\newcommand{\lftwo}{{\sc lat}field{\sc 2}} %

\newcommand{\nerd}[1]{{\small \tt #1}}

\title{\lftwo: a C++ library for classical lattice field theory}

\author{David Daverio\footnotemark[2]\ \footnotemark[3]  \and Mark Hindmarsh\footnotemark[4] \and Neil Bevis\footnotemark[4] }

\begin{document}
\maketitle

\renewcommand{\thefootnote}{\fnsymbol{footnote}}
\footnotetext[2]{D\'epartment de Physique Th\'eorique, Universit\'e de Gen\`eve, quai Ernest Ansermet 24, 1211  Gen\`eve  4, 	Switzerland}
\footnotetext[3]{david.daverio@gmail.com}
\footnotetext[4]{Department of Physics \& Astronomy, University of Sussex, Brighton BN1 9QH, U.K. }

\renewcommand{\thefootnote}{\arabic{footnote}}

\pagestyle{myheadings}
\thispagestyle{plain}
\markboth{D. Daverio, M. Hindmarsh and N. Bevis}{\lftwo}

\begin{abstract}

\lftwo\ is a C++  library designed to simplify writing parallel codes for solving partial differential equations, developed for application to classical field theories in particle physics and cosmology. It is a significant rewrite of the \lf\ framework, moving from a slab domain decomposition 
to a rod decomposition, where the last two dimension of the lattice are scattered into a two dimensional process grid. Parallelism is implemented using the Message Passing Interface (MPI) standard, and hidden in the basic objects of grid-based simulations: Lattice, Site and Field. It comes with an integrated parallel fast Fourier transform, and I/O server class permitting computation to continue during the writing of large files to disk.
\lftwo\ has been used for production runs on tens of thousands of processor elements, and is expected to be scalable to hundreds of thousands.
\end{abstract}

\section{Introduction}

The exponential increase in parallel computing power 
has meant that it  has become feasible to tackle ever larger computational problems.  However, the difficulty of using parallel machines has not decreased at the same rate.  In this paper we introduce a new package \lftwo\ designed to simplify numerical simulation on cartesian grids in parallel.  The distinguishing features of \lftwo\ are simplicity, scalability, and an integrated portable parallel fast Fourier transform (FFT).
Parallelism is implemented using Message Passing Interface (MPI) standard and can handle an arbitrary number of processes greater than or equal to 4, with grids of dimension 2 or higher. \lftwo\ has been used for the numerical solution of partial differential equations on 3D rectangular grids with production runs on over 30k processors.  

The package is a fork from the LATfield project, which was itself inspired by Matrix Distributed Processing (MDP) library \cite{DiPierro:2000bd,DiPierro:2003fm,DiPierro:2005jd}. MDP, now distributed as part of the larger FermiQCD library  \cite{DiPierro:2005qx}, is a C++ library for fast development of parallel distributed computations in lattice gauge theory. Like MDP, LATfield turns the basic concepts of a lattice simulation into the elementary objects: lattice (or grid), site and field; and also shares MDP's convenient site-referencing syntax. However, LATfield was developed from scratch to solve problems of a different class, involving larger grids and therefore needing improved memory efficiency, although at the cost of losing the topological versatility of MDP.

\lf\ was originally written to simulate cosmic strings in the 3-dimensional Abelian Higgs model \cite{Bevis:2006mj}, and has since been used for calculations over a wide range of subjects including MOND (Modified Newtonian Dynamics) \cite{Bevis:2009et} and Q-balls \cite{Tsumagari:2009na}. The machine available for the original work was  a supercomputer with 128 CPUs, which by today's standards would be considered very small, and therefore the software design is not necessarily appropriate for the some of the machines available today. Furthermore, that project used a public domain parallel Fast Fourier Transform (FFT ) package \cite{FFTW97}, which use slab domain decomposition: the cuboidal simulation volume is cut along one dimension such that each CPU is responsible for a slice of the total simulation volume. In  \lf, the domain decomposition of all data followed suit: hence the 1-dimensional parallelism.

\lftwo\ was written to overcome the limitation that slab domain decomposition yields: a cubic $N^3$ lattice can be parallelised over at most $N$ cores, and in practice slabs that are only a few sites thick are inefficient, since the communication overhead becomes signifiant. With ever larger supercomputers being available, this limitation became increasingly relevant. Therefore \lftwo\ was designed with rod-based parallelism, i.e. the lattice is cut along two dimensions. Indeed, the``2" in \lftwo\ refers to this number. Further, \lftwo\ incorporates a 3-dimensional FFT that shares this rod-based parallelism and that is also highly portable. Finally, \lftwo\ introduces an I/O server to allow computation to continue during disk writes. These modifications have enabled \lftwo\ to be used to study $4096^3$ lattices across over 32k cores.

In this article we introduce the library in the hope that it may be useful for others working on similar problems. \lftwo\ is very well suited to the numerical simulation of classical field equations in two or more dimensions. The library is extremely easy to use and therefore allows a very fast way to produce scalable code. The FFT wrapper allows the use of spectral methods and the spectral analysis of solutions. In this paper, we outline the main features and then present benchmarks of the library. In addition, examples can be found at the download page and are commented line by line in the library documentation.


Like its predecessor, the \lftwo\ library has been used to study cosmic strings via the classical field equations of the Abelian Higgs model  \cite{stringists:2014}. More recently \lftwo\ has been used to develop solver for General Relativity in the weak field limit \cite{Adamek:2013wja}. 


There are already a number of libraries for parallel simulation on grids, such as PetSc \cite{petsc-web-page}, Cactus \cite{Goodale2002a}, and CHOMBO \cite{chombo:xxx}. 
Such frameworks are much more complicated and have many more features: for example, \lftwo\ does not have adaptive mesh refinement (AMR) and does not contain any solvers. By constrast, \lftwo\ has been designed with simplicity in mind, for the rapid development of scalable parallel code on static grids.

A unique feature is a fast Fourier transform with 2-dimensional domain decomposition, running on the same cores as the simulation. This enables efficient spectral analysis for very large (currently only 3-dimensional) grids. 
\lftwo\ is therefore a complementary tool for field-based simulations, and provides users a way to build scalable executable with little previous experience of parallel computing.


\section{Getting started}

The website of \lftwo\ is \url{www.latfield.org} and the library is available on a public repository\footnote{\url{http://github.com/daverio/LATfield2}}. 
The package contains the library, a PDF version of the documentation, examples and benchmarks. The examples are the best starting point, as all important feature of \lftwo\ are presented and explained line by line within the documentation. There are three basic examples, explaining how to work with \lftwo\ basics, how use the FFT wrapper, and how to use the I/O server. In addition, there is a small Poisson solver which exhibits a more realistic usage of \lftwo.

\subsection{Compilation}

In its basic form, \lftwo\ should be compilable with any recent C++ compiler, without external libraries. If parallel I/O and FFTs are desired, then the HDF5 and FFTW3 libraries should be installed, and certain compilation flags set. The compilation flags are presented in table \ref{table:nref_alt} are used to compile the library with HDF5 (with optional PIXIE dataset format), the I/O server, and the fast Fourier transform.  Each library can be selected independently.

\begin{table}[htdp]
\centering
\caption{Compilation flags}
\begin{tabular}{| r | l |}
\hline
\multicolumn{2}{|c|}{\bf Fast Fourier transform flag}  \\
\hline 
\tt  FFT3D & enable FFT functionality  \\
\hline 
\hline 
\multicolumn{2}{|c|}{\bf HDF5 flags}  \\
\hline
\tt HDF5 &  enable HDF5 functionality\\
\hline
\tt  H5\_HAVE\_PARALLEL & enable parallel HDF5 (needs flag HDF5)  \\
\hline
\tt  H5\_HAVE\_PIXIE& use Pixie dataset format (needs flag HDF5) \\
\hline 
\hline 
\multicolumn{2}{|c|}{\bf IO Server}  \\
\hline 
\tt  EXTERNAL\_IO & enable the IO server \\
\hline 
\end{tabular}
\label{table:nref_alt}
\end{table}%

Note that the library cannot be pre-compiled, as it uses C++ templates. 
This slows down the compilation, but does not create major issues as a modern compiler should not take more than one minute to compile \lftwo. 

\section{Library structure}

\lftwo\ is based on four C++ classes: Parallel, Lattice, Field, Site, which have been developed to hide as much as possible of the parallelization from the user, and with the finite difference solution of partial differential equations on large static Cartesian grids in mind. 


Parallel computation needs communication between processes and such methods are defined within the Parallel object. This object stores the geometry of the MPI processes and has methods to communicate between processes and also to perform global operations such as min (find the minimal value of a variable through all processes).  
The description of the geometry of the mesh is encapsulated in the Lattice class. \lftwo\ works on cartesian meshes with a dimensionality bigger or equal to two, and currently implements only periodic boundary conditions. 
The template class Field is used to declare fields on the mesh stored within a Lattice object. The Field class stores the datatype of the field, a pointer to the Lattice object on which the field exists and a pointer to the field data array. 
The Site class is used to refer to the value of fields on a given lattice site. The Site class also has methods to address the neighbours of a site, as will as a method to scan through all sites.

\subsection{Parallel infrastructure}
\label{section::parallel}

\lftwo\ distributes $n$-dimensional cartesian lat\-tices onto a  2-dimensional cartesian grid of MPI processes, in a ``rod'' decomposition. The last dimension of the lattice is scattered into the first dimension of the process grid and the last-but-one dimension is scattered into the second dimension of the process grid. This choice has been made to increase data locality of the ghost cells (halo), which increases the efficiency of the method to update them. Due to its scheme of parallelization, \lftwo\ is only able to work with lattice of dimension bigger or equal to two.

The geometry of the process grid (the sizes of the two dimensions), two layers of MPI communicator and simple communication methods are embedded in the parallel object, which is an instance of the class Parallel. This object is instantiated but not initialized within the library header, therefore user should never declare an instance of the Parallel class. but rather use directly its instance ``parallel".

The parallel object is not initialized in order to allow the user to set the process grid geometry. The geometry of the grid is set at initialization of the parallel object, by passing the size of each dimension of the process grid. This is performed using the following method:
\begin{verbatim}

     parallel.initialize(int n, int m);

\end{verbatim}
where $n$ is the size of the first dimension of the process grid and $m$ of the second, a $n*m$ geometry. The initialization of the parallel object should be the first execution of any package which use \lftwo\ as all its classes depend on this object and more important as MPI is initialized within the parallel object initialization. 

One should notice than a geometry of $n*m$ is not equal to $m*n$. Indeed $n*m$ means that the first dimension has size $n$ and the second size $m$,  so $m*n$ mean the inverse. Scalability and execution time depend on the geometry. Usually, it is most efficient to have a close to square geometry. In many case a square geometry of the process grid is not possible (for example, due to node allocation schemes on the cluster or to the problem sizes). In that case, it is always better to use a $n*m$ geometry with $n<m$. This is especially true when using the fast Fourier transform functionality.


\subsubsection{I/O server}

The \lftwo\ parallelization allows the usage of additional MPI processes reserved for writing to disks. 
The server has been developed to allow the user to write relatively small amounts of data very fast, where ``small'' means 5-10\% of the total physical memory on a node. Examples would be the positions of zeros or the local maxima of a field. The server has been implemented to maximize its performance, regardless of the structure of output files. The server writes data into several files sharing a root filename. The server processes are grouped together, reflecting typical cluster architecture of multicore nodes. The number of file parts is defined by the number of server process groups. The I/O server currently has no capability for input. 

The files have no structure at all, and are composed of the list of messages sent by the compute processes. This simplicity minimizes the amount of computation done by the server, in the goal to maximize the time the server is ready to receive messages from the compute nodes. Therefore the output files will need some post production treatment to recover the correct data structure. 

\begin{figure}
\centering
\includegraphics[scale=1.3]{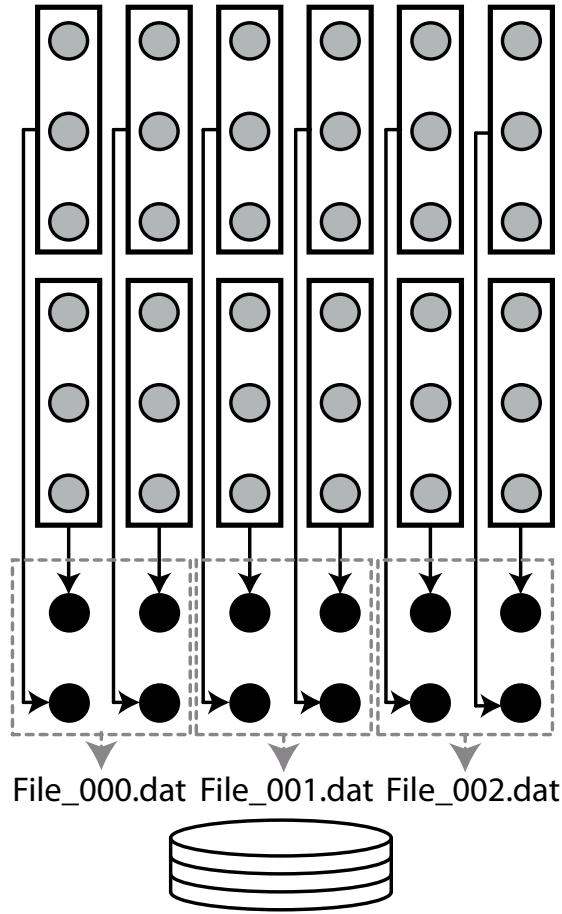}
\caption{\lftwo\ with its I/O server. 36 compute process in grey, 12 I/O process in black, 3 I/O group of 4 processes in grey dashed.  This geometry is set by the following command: parallel.initialize(6,6,12,4);  }
\label{f:ioserver}
\end{figure}

Figure \ref{f:ioserver}{} describes the geometry and the interconnection of the processes. Each server process is connected to several compute processes, and then combined into a group. Each server process group writes a single file using MPI parallel I/O. The server process group is best chosen to reside on a single node.

To use \lftwo\ with the I/O server the flag -DEXTERNAL\_IO must be set at compilation, and the initialization of the parallel object is done with the following method:
\begin{verbatim}

   parallel.initialize(int n,int m,int io_size,int io_group_size);

\end{verbatim}
where  {\verb n } and  {\verb m } have the same significance as before. 
The variable {\verb io_size } is the number of process allocated to the I/O server, which must be an integer devisor of  {\verb n }, and  {\verb n/io_size }  should be an integer divisor of  {\verb m }.  {\verb io_group_size } is the number of processes in each server group. It should be an integer divisor of  {\verb io_size }. 

The procedure to write data on disk with the IO server is the following. First an ostream is open by the compute processes. This ostream is the link between the compute and IO processes. It can be open only if the server is in the ready state, meaning that the server is started and is not writing data on disks. Therefore the procedure of opening an ostream returns a boolean to tell to the compute processes if the ostream is open. There is no wait-until-ready procedure: this needs to be hard coded by users. 

A successful opening of an ostream synchronises the compute and the IO processes. As the synchronisation implies communication over the entire network, there can be a non-negligible latency. However, with good balancing of IO and computation, there should always be a way to make this sync time negligible. 

Once an ostream is open, one can start to create files. Currently the server does not have not the capability to open a existing file: however, it can open and truncate existing files or create a new one. In addition, the current implementation can manage a maximum of 5 files per ostream, 
and only one file can be open at the same time. This issue will be solve in the next release of \lftwo, to allow all files to be open simultaneously. 

When all data have been transferred, and all files closed, the last procedure starts the transfer to disk by the server nodes. This is done when the compute nodes close the ostream. One should notice that from the moment the ostream is open to the moment that the server has finished writing data on disk the server is in the busy state. The busy period depends a lot on the architecture of the cluster: therefore it should be benchmarked to allow a good balancing. The procedure is explained in details in the IO server example of the documentation.



\subsection{Lattice}

\lftwo\ works on $n$-dimensional cartesian lattices, with $n$ bigger or equal to two. The Lattice class stores the description of the lattice geometry. As mentioned in section \ref{section::parallel}, the two last lattice dimensions are distributed over the processor grid. 
Therefore each processor will only have a subset of the lattice in memory; in \lftwo\ this part is referred as the local part. The local geometry description is also stored in the Lattice object. The local geometry is never set by the user, indeed the user can only choose the size of each lattice dimension and the halo size. \lftwo\ will then automatically scatter the lattice onto the processor grid. To allow working with multiple fields which require different halo sizes, the distribution over processes does not depend in any way on the halo size.

The following method will declare and initialize a lattice with dimension {\tt dim}, with a linear size in each dimension given by the array {\tt sizes}, which has {\tt dim} elements, and with a halo of {\tt h} sites. 
\begin{verbatim}

   Lattice lat(int dim, int * sizes, int h);

\end{verbatim}
One should notice that while the size of the halo is defined within the Lattice object, the {\tt updateHalo} method belongs to the Field class. This allows one to work with multiple fields on the same lattice, and to update the halos of the fields at different moments. 

\subsection{Field}

The class Field is a template class which describes data distributed over all the sites on a given lattice. The datatype of the field can be arbitrary, but field I/O is defined only for a list of given data type (see section \ref{subsection:IO}). The number of components of a field is set at initialization. The components can be arranged as a 1 dimensional array or a 2-dimensional matrix. In case of a matrix, one can specify that the matrix is symmetric to avoid redundant component. The last important variable of a field is the pointer to the data array. One should notice that this pointer does not need to be allocated when the Field object is instantiated.

The following method, will declare, initialize and allocate a field of datatype double on the lattice {\tt \small lat} with \nerd{comp} components.
\begin{verbatim}

   Field<double> phi(Lattice lat,int comp);

\end{verbatim}
Invoking this method is equivalent to the following 3 lines of code:
\begin{verbatim}

     Field<double> phi;
     phi.initialize(lat,comp);
     phi.alloc();

\end{verbatim}
Note that the initialize method does not allocate memory.

\subsubsection{Halo cells}

Halo cells or ghost cells are a layer of cells wrapping the local cells. These cells contain a copy of the data of the corresponding lattice cells inside the computational domain. The halo cells link neighbouring processes, but also encapsulate the periodicity of the lattice, which has the topology of a torus. The copy of the data into the halo is perform with the {\tt updateHalo()} method of the Field class. 
Notice that in \lftwo\ every dimension has a halo, even the local dimensions. This is done to increase locality of the data when working with neighbouring cells. 

\subsection{Site}
The site object stores the topology of the data distribution in the memory. The Site class gives to each local coordinate an index in the memory and vice versa. In \lftwo\ the Lattice class only store the size of each dimension, but has no description of how the data of a field is stored. Currently \lftwo\ contain only one type of site class which distribute the array in a row major order.

 A Site object is instantiated for a lattice \nerd{lat} with the following method:
\begin{verbatim}

     Site x(Lattice lat);

\end{verbatim}
Passing an instance of the Site class to a Field instance will return the value at the position pointed by the Site object. It is also used for assignment, as illustrated by the code segment
\begin{verbatim}

    double a = phi(x);
    phi(x)=1.0;

\end{verbatim}
where \nerd{phi} is assumed to have been initialised as type \nerd{double}. 
%

The Site class also contain methods to loop over the lattice: {\tt first()}, {\tt test()} and {\tt next()}. Which can be used within to build a loop as follows:
\begin{verbatim}

    for(x.first();x.test();x.next()){
       phi(x)=1.0;
    }

\end{verbatim}
This loop will travel in the memory allocated for the field by incrementing the index store by the Site class (performed by the \nerd{next()} method). As the data distribution is row major order, the indexing will goes from the 0 dimension to the last dimension.

Two important methods of the Site class are the overloaded operators $+$ and $-$.  Let us give an example in 3 dimension. If site {\tt x} refers to coordinate $(x,y,z)$ then:

\begin{quote}
{\tt x+0 } refers to $(x+1,y,z)$\\
{\tt x+2 } refers to $(x,y,z+1)$\\
{\tt x-0 } refers to $(x-1,y,z)$\\
{\tt x-1 } refers to $(x,y-1,z)$ 
\end{quote}

Hence, if one would like to form the symmetric difference of component \nerd{c} of a field \nerd{phi} at site \nerd{x} in the $x$ direction, we would write {\tt phi(x+0,c) - phi(x-0,c);}. This displacement in the lattice are perform by adding a ``jump" value at the index stored by the Site object. As the data distribution is row-order, the jump in the zero direction is equal to one, then for the first it is equal to the size of the zero dimension plus twice the halo size, and so on.

One should notice that ``shift'' operations as \nerd{phi(x+a) = phi(x)} or \nerd{phi(x-a) = phi(x)} are not defined in \lftwo.
Future versions will include a shift method.

If one wants to assign or work with a given coordinate one can use the method \nerd{setcoord} of the class Site. This method will set the site object to point to the given coordinate, which is passed as an array of length equal to the number of dimensions of the lattice. One should notice that this site of the lattice exist only in one MPI process. Only this process will have its site set to the coord. To select the correct process the method \nerd{setCoord} returns a boolean flag, which is true only if the coord is local. Therefore this method should be inside a conditional statement as:
\begin{verbatim}
     if(x.setCoord(r))phi(x)=1.0;
\end{verbatim}
with \nerd{r} an array of integers of length equal to the number of dimensions of the lattice.

\subsection{Input and output}
\label{subsection:IO}

Input and output methods, both serial and parallel, are provided with the field class.
%
%
The serial methods are inherited from {\sc lat}field. The \nerd{save} and \nerd{load} methods will respectively write and read a field in ASCII format in serial, with each process  taking turns to append to a file created (if necessary) by the root process.   
These I/O methods are very slow and produce large files: however they can be very useful at early stage of development as ASCII files are well suited for humans. The \nerd{write} and \nerd{read} methods are equivalent but perform I/O in binary data format. 

Both write methods will generate one single file, written independently of the processes grid geometry. The file will have no header 
and is basically a list of values in row-major order. In the ASCII version, each element of the array are separated by a new line character. For both binary and ASCII I/O, each element of this array is a list of the components of the field. 

\subsubsection{HDF5}

One major I/O improvement of \lftwo\ with respect to the first version \lf\ is the usage of HDF5. We have chosen HDF5 because it is a widespread standard library for High Performance Computing, and the format is embedded in python (h5py), matlab, mathematica and VisIt. This allows post production data analysis of large dataset without difficulties. 
In order to use HDF5, the compilation flag -DHDF5 should be set. This enables the methods \nerd{saveHDF5} and \nerd{loadHDF5}

By default \lftwo\ use the serial version of HDF5. To use the parallel version,  the flag -DH5\_HAVE\_PARALLEL has to be set at compilation. 

All native datatypes are allowed by the HDF5 wrapper of \lftwo, as well as the \lftwo-supplied class Imag for complex arithmetic. Multiple component fields of those datatypes are also supported.

One last comment on the HDF5 dataset format. The default setting of \lftwo\ will write all components of a multiple component field in a single dataset. In this way, the data structure of the dataset is very close to the one stored in the memory, which increases the efficiency of the write/read methods of HDF5 library. This choice is not very well suited to the HDF5 reader encapsulated in packages like VisIt. To ease the visualization and post production analysis, a second format can by used, based on the Pixie reader. In this format each component will be written in a separate dataset. This setting is set at compilation, using the flag -DH5\_HAVE\_PIXIE

\subsection{Fast Fourier Transform}

\lftwo\  contains a fast Fourier transform (FFT) algorithm based on the 1d serial FFT suppled with the FFTW package \cite{FFTW05}. To use it, FFTW version 2.2 or higher needs to be installed. The \lftwo\ FFT was developed for a classical field theory code LAH, and currently lacks some versatility. For example, the current version works only for 3d cubic lattices. Furthermore, the size of the third dimension has to be an integer multiple of both of the other dimensions of the processor grid. This limits the number of MPI processes which can be used for a given problem, but has been used with over 30k  MPI process with very good scalability, and we expect this scaling to extend to hundreds of thousand of processes. To enable the FFT capabilities of \lftwo\ the flag -DFFT3D must be set at compilation.

The FFT of \lftwo\ borrows the idea of a plan from FFTW. The FFT is handled by the class PlanFFT. This class creates a link between two instances of Field which are the ``same" field represented in Fourier and real space. To minimize the amount of memory used the PlanFFT class performs the allocation of memory for both instances. The transformation can be set to be in place or out of place, with  the default being out of place. When the transformation is performed in place, when the Fourier and real space representions of the field are instantiated they both point to the same data array. Some temporary memory is also allocated, as required by FFTW. This temporary memory is allocated within a variable which is not contained in the PlanFFT class, allowing multiple PlanFFT instance to share the same temporary memory. In addition the Fourier space lattice is initialized using the method from the Lattice class itself. Therefore a Fourier space lattice is defined from a real space lattice, which ensures that the Fourier space lattice has the correct size. The ``Poisson solver'' example demonstrates the usage of \lftwo\ FFT.

One important note is that the FFT does not preserve the data distribution. In the case of real to complex, if in real space dimension are labeled $(x,y,z)$ then in k-space they are $(k_x,k_z.k_y)$. For complex to complex, $(x,y,z)$ is transposed to $(k_z,k_x,k,_y)$. Therefore two additional Site classes are provided. One (rKSite) to work with the data distribution in Fourier space for a real-to-complex transform and the other one (cKSite) for a complex-to-complex transform.

\section{Benchmarks}

The library has been benchmarked on Monte Rosa (Cray XE6) at CSCS, the Swiss National Supercomputing Centre. The benchmark monitors the following parts of the library: 

\begin{remunerate}
\item FFT: complex to complex and real to complex transforms, both forward and backward.
\item Field operations: field assignment and referencing of fields at neighbouring sites by the construction of a difference operator.
\item Communication: the {\tt updateHalo} method is benchmarked for different size halos.
\item I/O: writing files using HDF5 parallel. 
\item I/O server: writing files using the server. 
\end{remunerate}


For all benchmarks we show the efficiency defined as follow:
\begin{equation}
E= \frac{n_{\rm ref}t_{\rm ref}}{n t_n},
\end{equation}
where $n$ is the number of MPI process of the run, $t_{n}$ the execution time with $n$ MPI process, $n_{\rm ref}$ the number of processes of the reference run, and $t_{\rm ref}$ the execution time of the reference run. Usually the reference run should be perform serially. However, as \lftwo\ has no capability to run serially, and as the required memory for one run is too large to be able to run the tests on one node for large lattice sizes, we have chosen $n_{\rm ref}$ as shown in table \ref{table:nref}. For both lattice sizes $512^3$ and $1024^3$ we have perform the benchmark for different numbers of field components, namely 1, 2, 3 and 6. For larger lattices, meaning $2048^3$ and $4096^3$, the benchmarks have been performed only with 1-component fields.

\begin{table}[htdp]
\begin{center}

\begin{tabular}{|c|c|}
\hline
Lattice Size & $n_{\rm ref}$ \\
\hline
$512^3$ & $4$ \\
\hline
$1024^3$ & $32$ \\
\hline
$2048^3$ & $32$ \\
\hline 
$4096^3$ & $128$ \\
\hline 

\end{tabular}
\caption{Number of MPI processes for the reference runs}
\end{center}
\label{table:nref}
\end{table}%

\begin{figure}[htbp]
\centering
\includegraphics[scale=0.8]{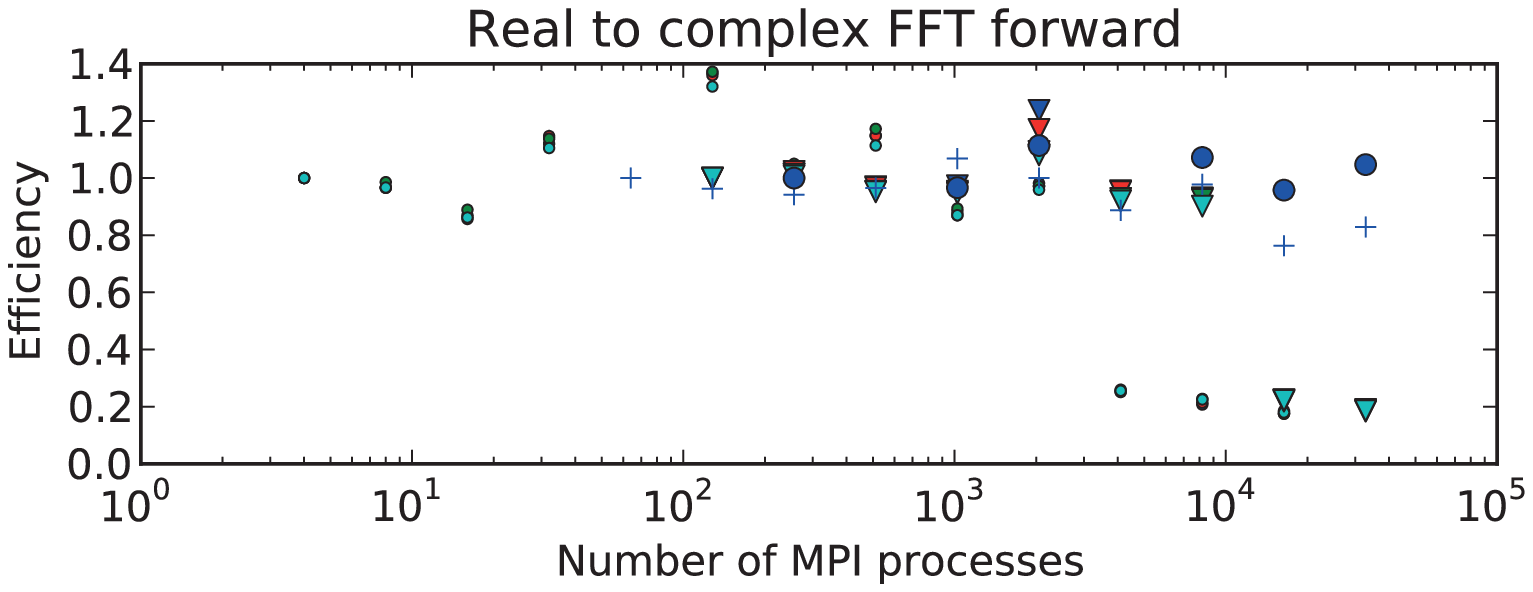}
\includegraphics[scale=0.8]{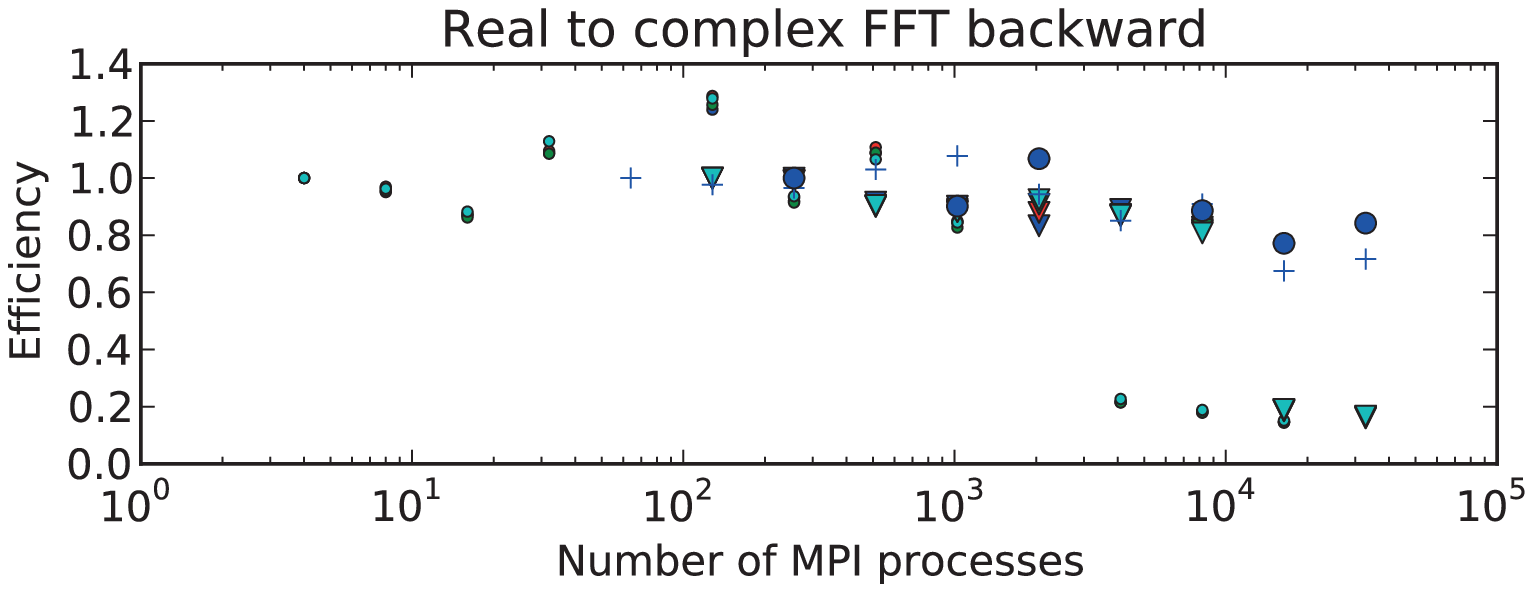}
\includegraphics[scale=0.8]{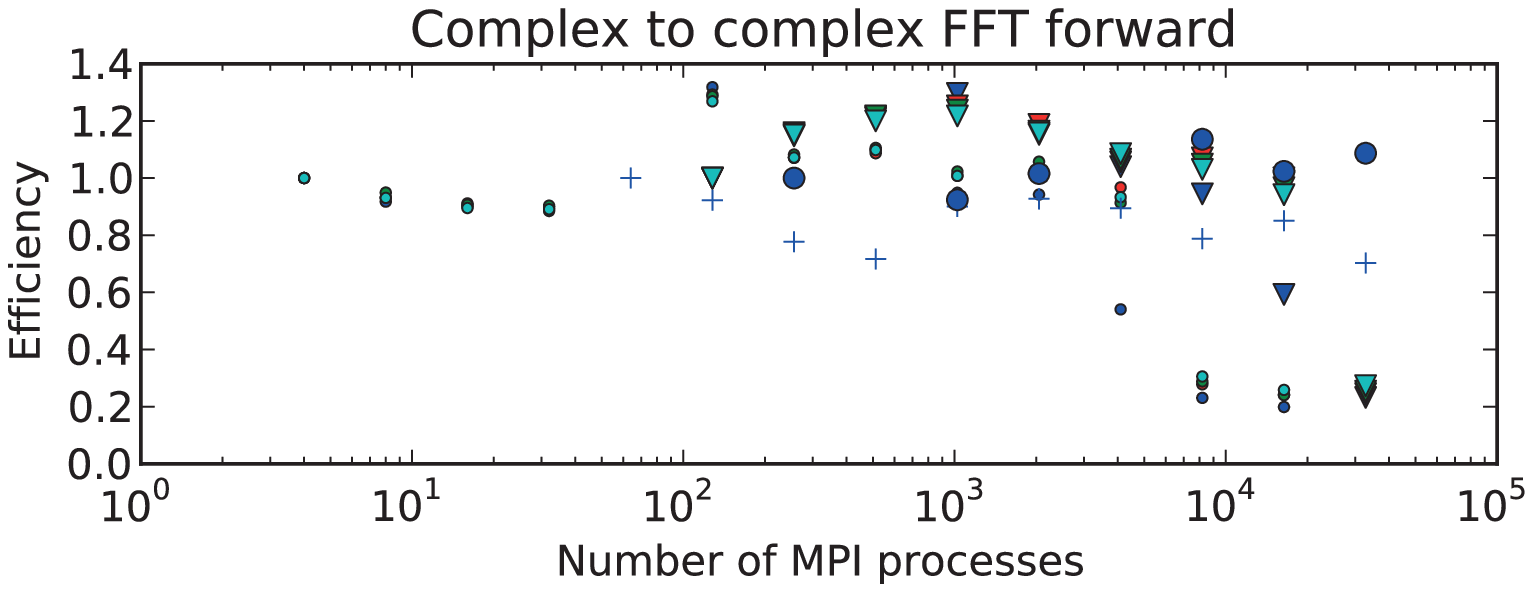}
\includegraphics[scale=0.8]{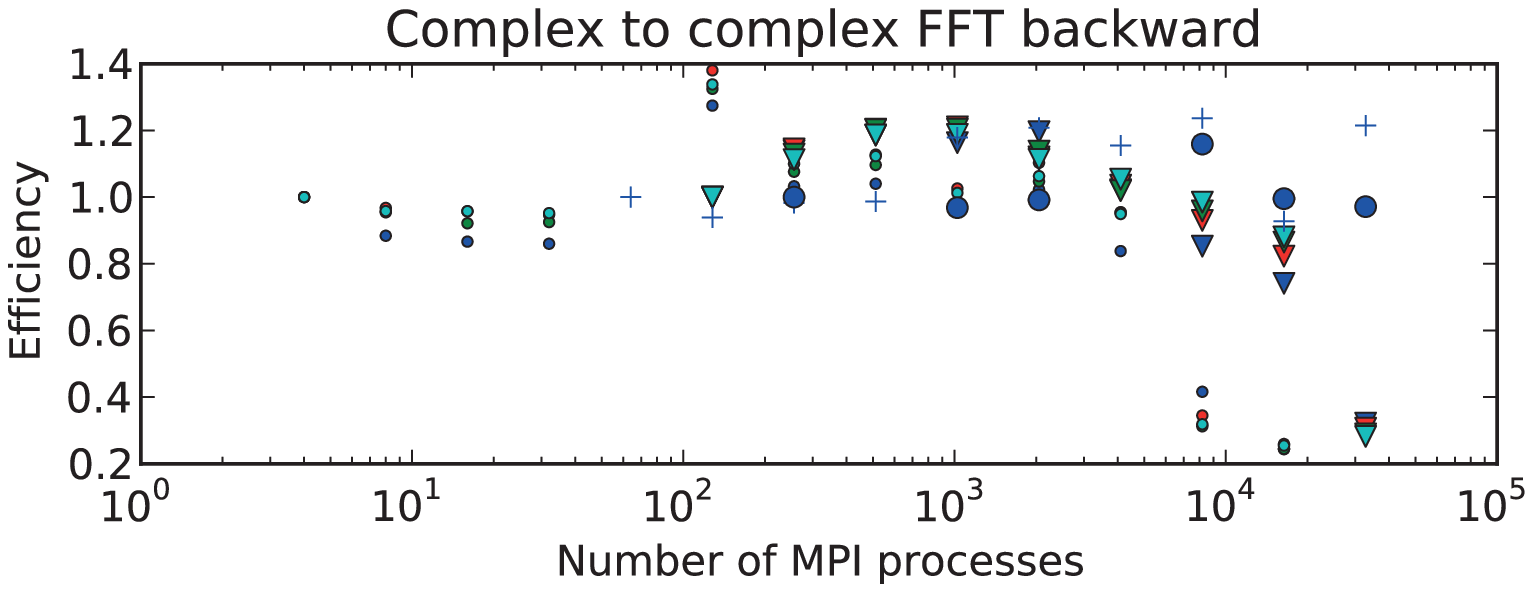}
\caption{ Fast Fourier Transform benchmarks. Dots: $512^3$ lattice sites, triangles: $1024^3$ sites, cross: $2048^3$ sites, circle: $4096^3$ sites. Blue: 1 field component, red: 2 components, green: 3 components, cyan: 6 components}
\label{f:bench_fourier}
\end{figure}

\begin{figure}[htbp]
\centering
\includegraphics[scale=0.78]{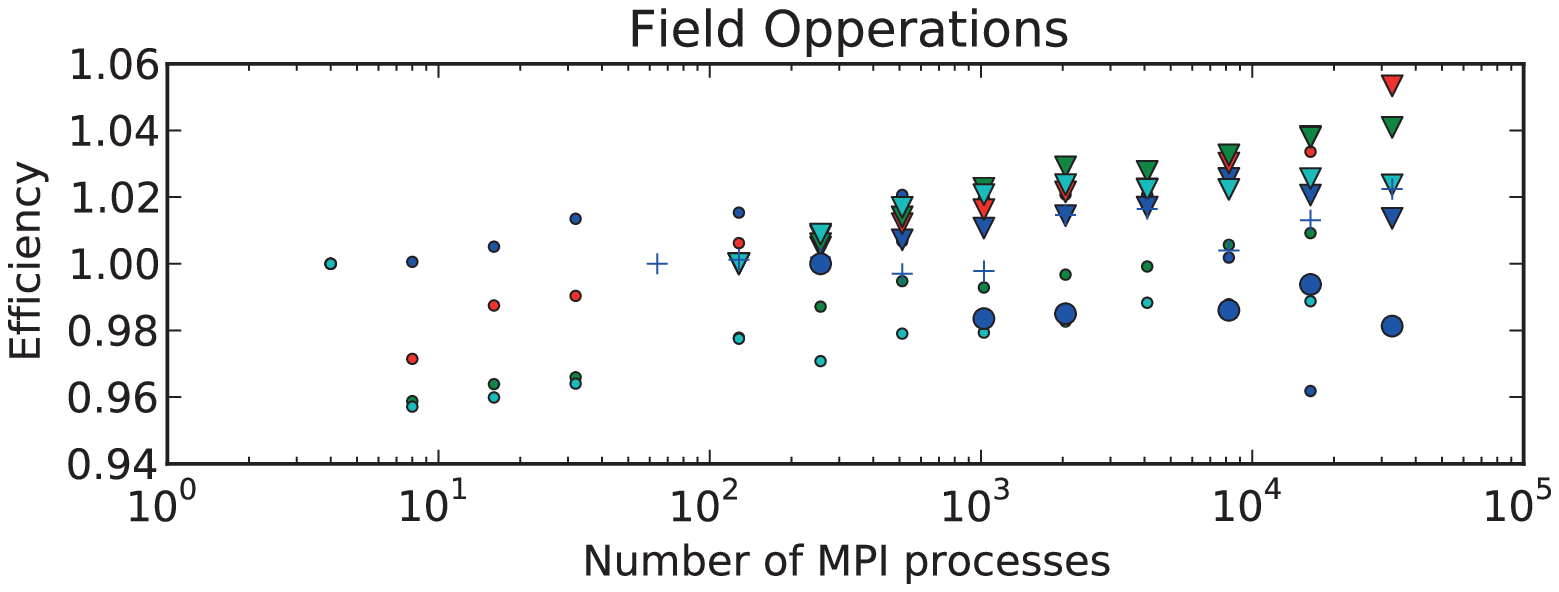}
\includegraphics[scale=0.8]{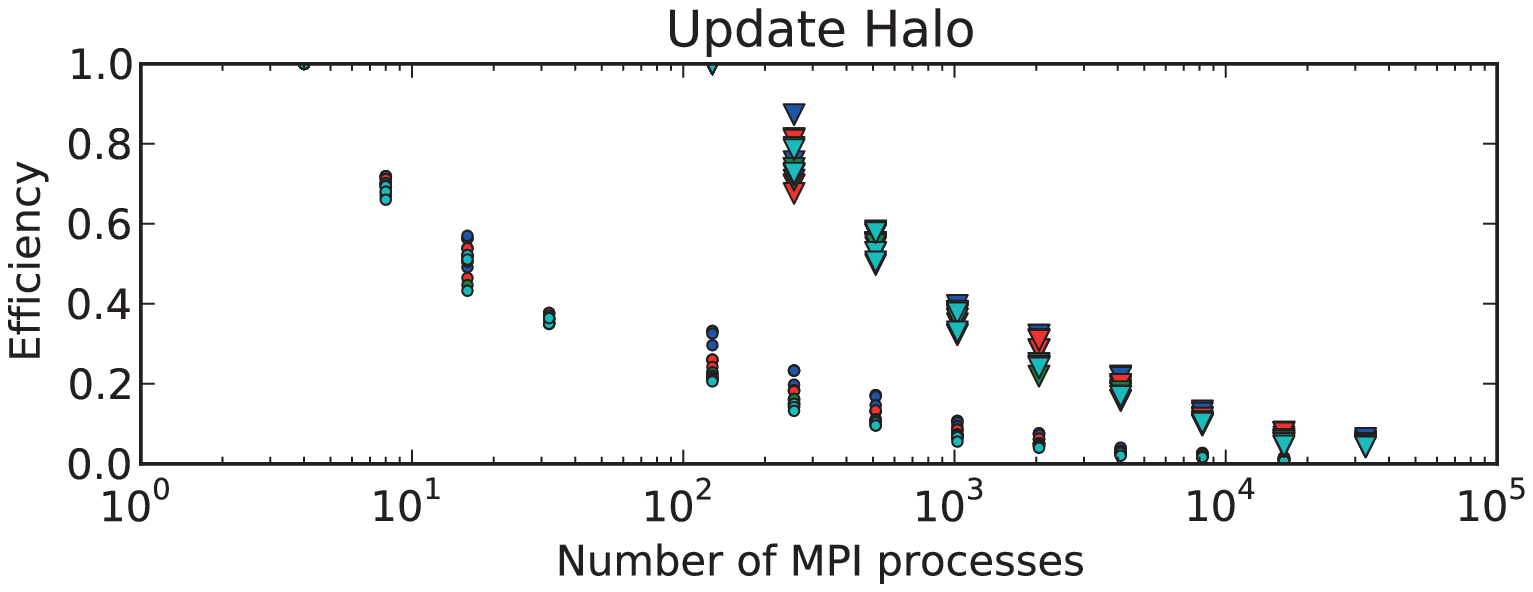}
\caption{ Field operations: local operation on fields. In that case computation of a first order discreet derivative. Update halo: procedure to update halo (ghost) cells. 
Dots: $512^3$ lattice sites, triangles: $1024^3$ sites, cross: $2048^3$ sites, circle: $4096^3$ sites. Blue: 1 field component, red: 2 components, green: 3 components, cyan: 6 components }
\label{f:bench_fopp}
\end{figure}

\begin{figure}[htbp]
\centering
\includegraphics[scale=0.7]{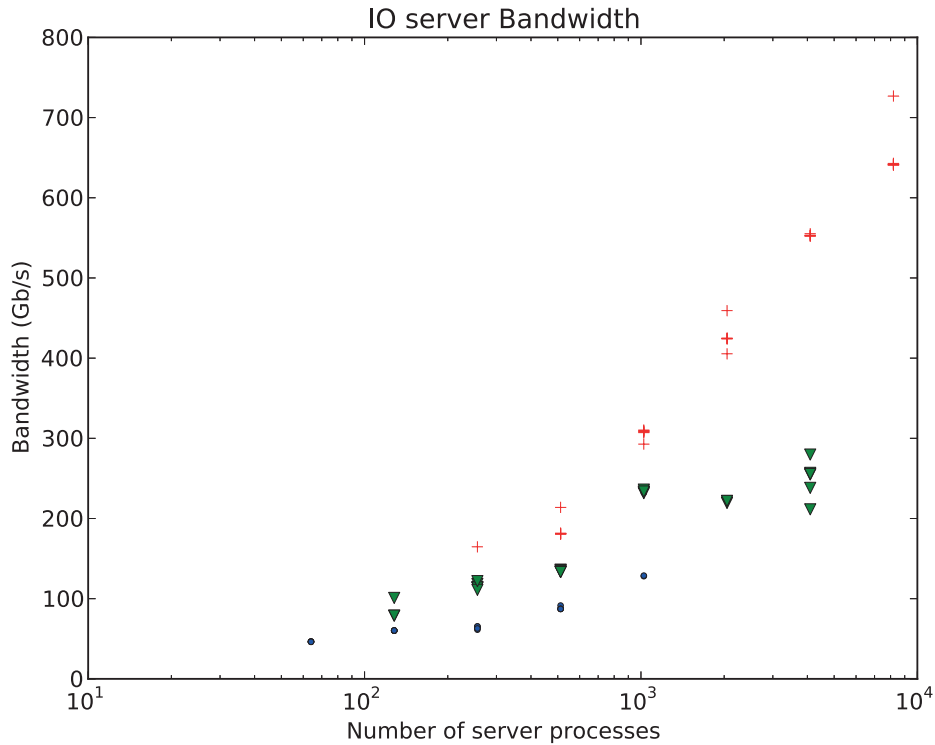}
\caption{Compute process to I/O Server bandwidth. Blue dot: 1024 compute processes, file size: 40GB; green triangle: 4096 compute processes, file size 80GB; red cross: 16384 compute processes, file size: 160GB. The fluctuation of bandwidth are mainly due to the fact that the server was in a busy state when the stream was open and therefor the sync time was large during benchmarking}
\label{f:bench_server}
\end{figure}

\begin{figure}[htbp]
\centering
\includegraphics[scale=0.7]{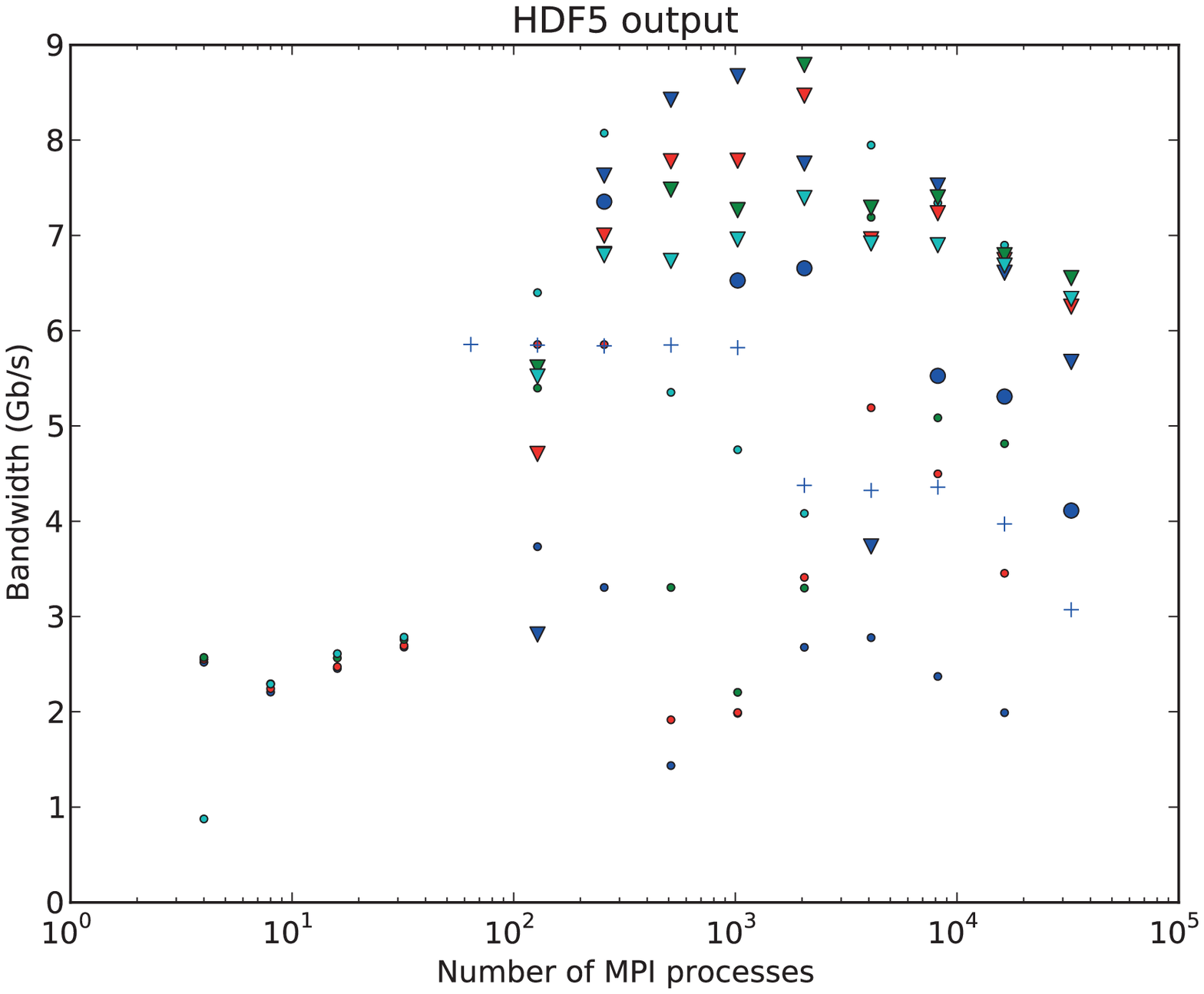}
\caption{ HDF5 output bandwidth. Dots: $512^3$ lattice sites, triangles: $1024^3$ sites, cross: $2048^3$ sites, circle: $4096^3$ sites. Blue: 1 component, red: 2 components, green: 3 components, cyan: 6 components }
\label{f:bench_hdf5}
\end{figure}

\subsection{Fast Fourier Transform}

The benchmarks of the Fast Fourier wrapper presented on figure \ref{f:bench_fourier} shows excellent scalability.  Indeed, one can use up to 4096 processes for lattices of $512^3$ sites with an efficiency close to 1 with respect to the 4-core times. After 4096 processes the efficiency drops and there is less gain in using larger number of processes. One can see that this drop is dependent on the lattice size. Indeed, for lattices of $1024^3$ sites the drop appears with 16384 processes. We expect to see this drop also for bigger lattices, but this has not been benchmarked, due to the lack of a larger computer. The drop is expect to be at 65536 process for a $2048^3$ lattice and at 262144 processes for $4096^3$ lattices. One can also notice that the wrapper efficiency depends only slightly on the number of components, but shows that a one component field is preferred.

\subsection{Field operations and update halo}
We test the field operations by the computation of the first order spatial derivative of all components followed by storing the result in an additional field. This operation does not involve communication between the processes, it is a purely local operation. Therefore this operation should have an efficiency close to 1. Figure  \ref{f:bench_fopp} shows that the efficiency is growing with the number of processes. This behavior is mainly due to an improvement of the data locality when using larger number of processes.

The update halo procedure does not scale well,  
as the size of the halo does not decrease as fast as the local part of the lattice as the number of processes is increased. However, this is not an issue in practice, and the total updateHalo procedure execution time is usually negligible compared to the compute time for local operations. 

\subsection{Outputs}

The  default settings of HDF5 outputs have been benchmarked and result are shown in Figure \ref{f:bench_hdf5}. It has been tested for a field with 1 component on a $512^3$ lattice (1 GB) to a field of 1 component on a $4096^3$ lattice (0.5 TB). Result show a large scatter of the bandwidth, this is not something reflecting any feature of \lftwo. The scatter is mainly due to the fact that the cluster used was not reserved during the benchmark, therefore other users could be using the disk at the benchmark time, which will scatter the bandwidth between users. One can notice that the maximum bandwidth is 8.8 GB/s, close to the maximum bandwidth of the Monte Rosa cluster (about 10 GB/s).

\subsubsection{IO Server}

The server has been benchmarked with 1024, 4096 and 16384 compute processes. We have test the server with a number of IO processes up to the number of compute processes. The benchmark shown on Figure \ref{f:bench_server}  include the synchronization and the transfer of data. With 1024 compute processes a total of 40 GB have been transferred, for 4096 compute processes 80 GB and for 16384 compute processes 160 GB.

\section{Conclusions}

We have introduced \lftwo\ \cite{latfield-web-page}, a simple C++ library for parallel numerical solution of partial differential equations on $n$-dimensional grids. The structure of the library has been described. The benchmark of the main feature of \lftwo\ have been discussed, showing an excellent scalability on distributed memory system as the Cray XE6.  \lftwo\ has been use in a real application (LAH, \cite{stringists:2014}) during 3 years of production at the Swiss National Supercomputing Centre (CSCS) running on  over 34k cores.

Future developments include a release of a Fast Fourier Transform working on $n$-dimensional lattice $(n>3)$ and for rectangular lattices,  a fast Fourier transform wrapper for accelerators using openACC and cuFFT, and a particle class including particle-to-mesh methods. The particle class will allow simulations with particles as well as fields, giving \lftwo\ users the capability to simulate (for example) gravitational clustering. 

\paragraph{Acknowledgments}

We thank Kari Rummukainen, Martin Kunz and Jon Urrestilla for comments on the manuscript. 
This work was supported by three grants from the Swiss National Supercomputing Centre (CSCS) under projects ID s319 and s546.
MDP is copyrighted by MetaCryption, LLC and it is released for free for non commercial research applications. 

\bibliographystyle{siam}
\bibliography{latfield2}

\begin{thebibliography}{10}

\bibitem{Adamek:2013wja}
{\sc Julian Adamek, David Daverio, Ruth Durrer, and Martin Kunz}, {\em {General
  Relativistic $N$-body simulations in the weak field limit}}, Phys.Rev., D88
  (2013), p.~103527.

\bibitem{chombo:xxx}
{\sc M.~Adams, P.~Colella, D.~T. Graves, J.N. Johnson, N.D. Keen, T.~J.
  Ligocki, D.~F. Martin, P.W. McCorquodale, D.~Modiano, P.O. Schwartz, T.D
  Sternberg, and B.~Van Straalen}, {\em Chombo software package for amr
  applications}, Design Document, Lawrence Berkeley National Laboratory
  Technical Report LBNL-6616E.

\bibitem{petsc-web-page}
{\sc Satish Balay, Shrirang Abhyankar, Mark~F. Adams, Jed Brown, Peter Brune,
  Kris Buschelman, Lisandro Dalcin, Victor Eijkhout, William~D. Gropp, Dinesh
  Kaushik, Matthew~G. Knepley, Lois~Curfman McInnes, Karl Rupp, Barry~F. Smith,
  Stefano Zampini, and Hong Zhang}, {\em {PETS}c {W}eb page}.
\newblock \url{http://www.mcs.anl.gov/petsc}, 2014.

\bibitem{Bevis:2006mj}
{\sc Neil Bevis, Mark Hindmarsh, Martin Kunz, and Jon Urrestilla}, {\em {CMB
  power spectrum contribution from cosmic strings using field-evolution
  simulations of the Abelian Higgs model}}, Phys.Rev., D75 (2007), p.~065015.

\bibitem{Bevis:2009et}
{\sc Neil Bevis, Joao Magueijo, Christian Trenkel, and Steve Kemble}, {\em
  {MONDian three-body predictions for LISA Pathfinder}}, Class. Quant. Grav.,
  27 (2010), p.~215014.

\bibitem{latfield-web-page}
{\sc David Daverio, Mark Hindmarsh, and Neil Bevis}, {\em {LAT}field {W}eb
  page}, 2015.
\newblock http://www.latfield.org/.

\bibitem{stringists:2014}
{\sc David Daverio, Mark Hindmarsh, Martin Kunz, Joanes Lizarraga, and Jon
  Urrestilla}, {\em {Energy-momentum correlations for Abelian Higgs cosmic
  strings}}, In preparation.

\bibitem{DiPierro:2000bd}
{\sc Massimo Di~Pierro}, {\em {Matrix Distributed Processing: A set of C++
  Tools for implementing generic lattice computations on parallel systems}},
  Comput.Phys.Commun., 141 (2001), pp.~98--148.

\bibitem{DiPierro:2003fm}
\leavevmode\vrule height 2pt depth -1.6pt width 23pt, {\em {A Bird's eye view
  of matrix distributed processing}}, Lect.Notes Comput.Sci.,  (2003).

\bibitem{DiPierro:2005jd}
\leavevmode\vrule height 2pt depth -1.6pt width 23pt, {\em {Parallel
  programming with matrix distributed processing}},  (2005).

\bibitem{DiPierro:2005qx}
{\sc Massimo Di~Pierro and Jonthan~M. Flynn}, {\em {Lattice GFT with
  FermiQCD}}, PoS, LAT2005 (2006), p.~104.

\bibitem{FFTW97}
{\sc Matteo Frigo and Steven~G. Johnson}, {\em The fastest {Fourier} transform
  in the west},  (1997).

\bibitem{FFTW05}
\leavevmode\vrule height 2pt depth -1.6pt width 23pt, {\em The design and
  implementation of {FFTW3}}, Proceedings of the IEEE, 93 (2005), pp.~216--231.
\newblock Special issue on ``Program Generation, Optimization, and Platform
  Adaptation''.

\bibitem{Goodale2002a}
{\sc Tom Goodale, Gabrielle Allen, Gerd Lanfermann, Joan Mass{\'o}, Thomas
  Radke, Edward Seidel, and John Shalf}, {\em The {Cactus} framework and
  toolkit: Design and applications}, in Vector and Parallel Processing --
  VECPAR'2002, 5th International Conference, Lecture Notes in Computer Science,
  Berlin, 2003, Springer.

\bibitem{Tsumagari:2009na}
{\sc Mitsuo~I. Tsumagari}, {\em {Affleck-Dine dynamics, Q-ball formation and
  thermalisation}}, Phys. Rev., D80 (2009), p.~085010.

\end{thebibliography}

\end{document}

\fi